\documentclass[11pt]{article}
\usepackage{amsmath,amssymb,amscd}

\usepackage{cite}
\usepackage{amsmath, amsthm, amssymb, amsfonts}
\usepackage{dsfont}
\usepackage{float}
\usepackage[a4paper]{geometry}
\usepackage{color}

\definecolor{red}{rgb}{1,0,0}

\usepackage{graphicx}
\usepackage[mathscr]{eucal}

\large\normalsize

\title{The invariance and non-decreasing expectation of an\\ evolutionary path characteristic under weak selection}
\author{Yun-Yun Yu$^{1,\#}$, Cang Hui$^{2,3,4, \#}$ Tian-Jiao Feng$^{5,\#}$, Cong Li$^{6}$, Chao Wang$^{6}$,\\ Yi Tao$^{5,6}$\footnote{Author for correspondence, and e-mail: yitao@ioz.ac.cn} 
and Rui-Wu Wang$^{6,7}$\footnote{Author for correspondence, and e-mail: wangrw@nwpu.edu.cn}
\\
$^1$School of Mathematics and Statistics, Northwestern Polytechnical University,\\
Xi'an 710072, P.R. China \\
$^2$Department of Mathematical Sciences, Stellenbosch University,\\
Stellenbosch 7602, South Africa\\
$^3$National Institute for Theoretical and Computational Sciences,\\
African Institute for Mathematical Sciences,\\
Stellenbosch 7602, South Africa\\
$^4$International Initiative for Theoretical Ecology, London, N1 2EE, UK\\
$^5$Key Laboratory of Animal Ecology and Conservation Biology,\\
Center for Computational and Evolutionary Biology, \\
Institute of Zoology, Chinese Academy of Sciences, \\
Beijing 100101, P.R. China \\
$^6$School of Ecology and Environment, Northwestern Polytechnical University,\\
Xi'an 710072, P.R. China \\
$^7$College of Life Sciences, Zhejiang University, Hangzhou 310058, P.R. China\\
$^{\#}$These authors have the same contribution to this paper. }

\date{}

\begin{document}

\maketitle

\newpage

\section*{Abstract}

Fisher’s fundamental theorem of natural selection states that the rate of change in a population’s mean fitness equals its additive genetic variance in fitness. This implies that mean fitness should not decline in a constant environment, thereby positioning it as an indicator of evolutionary progression. However, this theorem has been shown to lack universality. Here, we derive the Fokker-Planck equation that describes the stochastic frequency dynamics of two phenotypes in a large population under weak selection and genetic drift, and develop a path integral formulation that characterizes the probability density of phenotypic frequency. Our formulation identifies that, under both selection and genetic drift, the ratio of the probability density of adaptive traits (e.g., phenotypic frequency) to that under neutrality represents a time-invariant evolutionary path characteristic. This ratio quantifies the cumulative effect of directional selection on the evolutionary process compared to genetic drift. Importantly, the expected value of this ratio does not decline over time. In the presence of fitness variance, the effect of directional selection on expected phenotypic changes accumulates over time, diverging progressively from paths shaped solely by genetic drift. The expectation of this time-invariant ratio thus offers a robust and informative alternative to mean fitness as a measure of progression in stochastic evolutionary dynamics.

\newpage
\emph{\textbf{Introduction.}} Natural selection favors traits that enhance individual fitness, leading to the proliferation of a population and thereby increasing the population's mean fitness over time \cite{Price1972a, Frank1998}. Fisher elaborated on evolution via Darwinian natural selection by proposing the "fundamental theorem of natural selection", which asserts that the rate at which the mean fitness of a population increases for any species is directly proportional to the genetic variance in fitness within the population \cite{fisher1930}. A refined interpretation, termed the mean fitness increase theorem (MFIT), posits that the rate of change in a population's mean fitness is always nonnegative and precisely equals the variance in fitness \cite{ewens2004, baez2021}. Thus, increase in mean fitness has come to be widely regarded as a principal measure of evolutionary progress, both in natural populations and experimental systems \cite{Orr2009,Lenski2017,Couce2024}.

As evolution is primarily driven by four processes (natural selection, genetic drift, migration, and mutation), the partial rate of evolutionary change in Fisher's theorem refers specifically to the contribution of natural selection due to differential reproduction of genotypes. This holds true only under the assumption of a constant environment, where the average effects of alleles on individual fitness remain fixed \cite{price1972b, ewens1989, edwards1994, Okasha2008, ewens2024}. In real populations, individual fitness is shaped by both phenotypic variation within the population and interactions with other species \cite{nowaksigmund2004, Hui2022}, while environmental variability adds another layer of complexity, as biological performance – and thus fitness – shifts across different environmental conditions \cite{Barton and Turelli1987, Schoener2011, Hendry2017}. As a result, individual fitness is inherently relative and often fluctuates over time, frequently resulting in a non-zero rate of change, which indicates that mean fitness does not always increase over evolutionary time scales \cite{Grafen2015}. Importantly, individual fitness can also be frequency-dependent, meaning that mean fitness is not strictly non-decreasing during evolution. Consequently, also as Fisher anticipated, there may exist a population-level characteristic more fundamental than mean fitness, one that remains non-decreasing under more general forms of natural selection.

Fisher’s theorem addresses only the deterministic process of natural selection through changes in genotypic frequencies, without explicitly accounting for genetic drift, which is pervasive in finite populations \cite{edwards1994,Wright1931}. Genetic drift can alter genotypic frequencies at equilibrium and shorten the fixation time, thereby modifying evolutionarily stable strategies in evolutionary game dynamics \cite{Liekens2004}. Subsequent analyses have underscored these limitations, fueling ongoing debates about the theorem’s relevance in evolutionary biology \cite{ewens1992, castilloux1995, lessard1995, lessard1997, tao2000, ewens2015}. A more comprehensive theoretical framework – one that accounts for both genetic drift and frequency-dependent selection, rather than adhering to the simplifying assumptions of Fisher’s original theorem – is essential for accurately modeling evolutionary processes in finite populations \cite{Poelwijk et al.2007, Traulsen2007, Gokhale2009}.

Here, we examine a broad model of evolutionary game theory developed by Maynard Smith in the 1980s, which incorporates both frequency dependence and demographic stochasticity \cite{maynardsmith1982}. By introducing stochastic dynamics into evolutionary games within finite populations under weak selection \cite{hofbauersigmund1998, traulsen2005, nowak2006}, we aim to derive a fundamental evolutionary characteristic using a path integral formulation. This formulation has been employed to explore fitness fluxes \cite{mustonen2010}, Muller's ratchet in population genetics \cite{Neher2012}, the probability density of the Wright-Fisher process under constant selection \cite{Schraiber2014}, and selection coefficients \cite{sohail2020}. The path integral formulation could serve as a powerful alternative methodology for modeling the stochastic evolutionary processes of both natural selection and genetic drift.

The stochastic model presented in this paper reveals two core features of evolutionary dynamics: the invariance of an evolutionary path characteristic and the expected evolutionary progression along realized evolutionary paths. Rather than relying on the traditional measure of mean fitness, the evolutionary path characteristic identified in this model offers a direct representation of the effects and dynamics of natural selection within stochastic evolutionary games. It offers a more accurate description of the processes of natural selection and genetic drift in shaping phenotypic frequencies, as opposed to Fisher’s theorem, which focuses solely on evolution by natural selection, or Kimura’s neutral evolution, which is influenced solely by genetic drift \cite{kimura1983}.

\emph{\textbf{Fisher’s mean fitness in evolutionary games.}} Following the standard framework of evolutionary game theory \cite{nowak2004, nowak2006, zheng2011}, we analyze a two-phenotype evolutionary game in a finite population of fixed size $N$, where two phenotypes (or strategies) are represented by $A$ and $B$. The payoff matrix is structured as
\begin{align}
M = \begin{pmatrix} a & b \\ c & d \end{pmatrix} \nonumber \ ,
\end{align}
where $a$ (or $b$) is the payoff for $A$ when competing against $A$ (or $B$), and $c$ (or $d$) is the payoff for $B$ against $A$ (or $B$). For convenience, we denote the alternating sum of the matrix as $als(M)=(a+d)-(b+c)$, and the determinant as $det(M)=ad-bc$. Let $x$ denote the frequency of $A$ in the population, so the frequency of $B$ is $1-x$. Hereafter, we discuss the short-term evolutionary dynamics when the frequencies have not reached the fixation states ($x=0,1$). Under random pairwise interactions in a large population size $N$, the expected payoffs for $A$ and $B$ are $\phi_A(x) = xa +(1-x)b$ and $\phi_B(x) = xc +(1-x)d$, with the mean payoff of the population given by $\bar{\phi}(x)= x\phi_A(x)+(1-x)\phi_B(x)$. The frequency $x^*=(d-b)/als(M)\in(0,1)$ corresponds to the strategy at which the two phenotypes receive equal expected payoffs, $\phi_A(x^*)=\phi_B(x^*)=det(M)/als(M)$, where $als(M) \ne 0$. This defines the unique evolutionary singularity in the two-phenotype game, where the selection gradient vanishes.

The fitness values of phenotypes $A$ and $B$, denoted by $f_A(x)$ and $f_B(x)$, are defined as $f_A(x)=(1-w)+w\phi_A(x)$ and $f_B(x)=(1-w)+w\phi_B(x)$, where $w$ represents the selection intensity ($w \in [0,1]$), indicating the extent to which the game influences individual fitness \cite{nowak2004, nowak2006, zheng2011}. Consequently, the mean fitness of the population is given by $\bar{f}(x)=xf_A(x)+(1-x)f_B(x)$. The time derivative of the mean fitness can be expressed as (see Sect. A of the appendix):
\begin{align}
\frac{d \bar{f}(x)}{dt}=\sigma_f^2(x)+x \frac{df_A(x)}{dt}+(1-x) \frac{df_B(x)}{dt},
\end{align}
where $\sigma_f^2 (x)=w^2x(1-x)(\phi_A(x)-\phi_B(x))^2$ represents the fitness variance of the population. This can also be expressed as $\sigma_f^2 (x)=w^2 x(1-x)als(M)^2 (x-x^* )^2$, suggesting $\sigma_f^2 (x)=0$ for all $x$ either under neutrality ($w=0$) or in games where $als(M)=0$. The fitness variance also vanishes when the phenotypic frequency is at either of the fixation states or the evolutionary singularity, i.e., $x=0,1,x^*$. This equation elucidates that the change in mean fitness comprises both the fitness variance ($\sigma_f^2 (x)$) and the average change in individual fitness ($x \frac{df_A(x)}{dt}+(1-x) \frac{df_B(x)}{dt}$).

When the game does not affect fitness (i.e., $w=0$), fitness becomes constant ($\frac{df_A(x)}{dt}=\frac{df_B(x)}{dt}=0$), and the mean fitness of the population cannot decline ($\frac{d\bar{f}}{dt}=\sigma_f^2 (x)=0$). In such a case, the mean fitness of a population will become non-decreasing, consistent with the MFIT. However, when the game does impact fitness ($w>0$), fitness becomes frequency-dependent and time-varying ($\frac{df_A(x)}{dt},\frac{df_B(x)}{dt}\ne0$), suggesting that the mean fitness of a population may not always increase (\textbf{Fig. A1}). As a result, mean fitness is no longer a monotonic indicator of evolutionary progression in such games. It also indicates that Fisher’s fundamental theorem holds only under its specific assumptions: deterministic evolution solely by natural selection, with constant and frequency-independent fitness. The theorem does not apply in evolutionary games where fitness is frequency dependent.

The deterministic, frequency-dependent evolutionary game described in Eq. (1) often fails to capture the contingent and stochastic nature of evolutionary dynamics \cite{Blount2018}. When genetic drift is considered, the evolutionary game dynamics can be better captured by a Moran process \cite{nowak2006}, where evolutionary trajectories become stochastic and irreproducible (\textbf{Fig. A2}). The statistical properties of these stochastic evolutionary paths remain largely uncharacterized. Questions such as whether some characteristics of these paths remain invariant over short time scales or indicate evolutionary progression, akin to the role of mean fitness in Fisher’s theorem, are still open and warrant systematic investigation. Addressing these questions requires both an alternative formulation of stochastic evolutionary game dynamics and a framework to quantify the characteristics of these stochastic paths. In what follows, we first derive a diffusion approximation of the stochastic dynamics, then present a path integral solution to the resulting evolutionary dynamics and finally identify characteristic features of these stochastic evolutionary paths that could potentially play a role analogous to mean fitness in Fisher’s theorem.

\textbf{\emph{The Fokker-Planck equation of evolution.}}
To develop an alternative evolutionary path characteristic and indicator of progression, we model the two-phenotype evolutionary game as a discrete-time Moran process \cite{nowak2006}. In each time step, an individual is chosen for reproduction with a probability proportional to its fitness, and its offspring, bearing the same phenotype, replaces a randomly selected individual within the population. Under the condition of a large population size $N$ \cite{traulsen2006, zheng2011}, the transition probabilities for the frequency of phenotype $A$ moving from $x$ to $x+1/N$ and from $x$ to $x-1/N$ are given by, $\pi^+(x)=x(1-x)f_A(x)/\bar{f}(x)$ and $\pi^-(x)=x(1-x)f_B(x)/\bar{f}(x)$, respectively. Consequently, $D^{(1)}(x)=\pi^+(x)-\pi^-(x)$ can be understood as the directional selection due to payoff differences between the two phenotypes. This term $D^{(1)}(x)$ becomes zero either under neutrality ($w=0$) or when the frequency is at either of the fixation states or the evolutionary singularity, i.e., $x=0,1,x^*$. Additionally, $D^{(2)}(x)=(\pi^+(x)+\pi^-(x)))/(2N)$ represents the per-capita genetic drift term in this stochastic evolutionary process. This term reduces to $x(1-x)/N$ under neutrality, which aligns with the variance in phenotypic frequencies per generation for a finite haploid population in the Wright-Fisher model.

Let $p(x,t;x_0)$ represent the probability density of strategy $A$ having frequency $x$ at time $t$, given that its initial frequency is $x_0 \in (0,1)$. In the diffusion approximation of the Moran process \cite{traulsen2006, zheng2011}, the Fokker-Planck equation for $p(x,t;x_0)$ can be derived via Taylor expansion as (see Sect. B of the appendix):
\begin{align}
\frac{\partial p(x,t;x_0)}{\partial t} &= -\frac{\partial}{\partial x} D^{(1)}(x) p(x,t;x_0) +\frac{\partial^2}{\partial x^2} D^{(2)}(x) p(x,t;x_0) \ .
\end{align}
This equation describes the time evolution of the probability density $p(x,t;x_0)$ due to the effects of selection ($D^{(1)}$) and genetic drift ($D^{(2)}$). Under neutrality ($w=0$), the equation simplifies to \cite{kimura1955}:
\begin{align}
\frac{\partial \tilde{p}(x,t;x_0)}{\partial t} &= \frac{\partial^2}{\partial x^2} \frac{x(1-x)}{N} \ \tilde{p}(x,t;x_0) \ ,
\end{align}
where $\tilde{p}(x,t;x_0)$ denotes the probability density of strategy $A$ under neutrality. The dynamics of $p(x,t;x_0)$ in Eq. (2) describes the evolution of the phenotype over time in a finite population. However, it is well known that the Fokker-Planck equation of evolution under frequency-dependent natural selection and genetic drift lacks a closed-form analytic solution. To uncover the evolutionary trajectory characteristics of stochastic evolutionary dynamics, we next derive the path integral formulation of the Fokker-Planck equation.

\textbf{\emph{Path integral formulation under weak selection.}}
Evolution can be envisioned as progressing along numerous probabilistic paths within the strategy space over time, and each path represented by a sequence of consecutive steps. For a specific evolutionary path, denoted as $\mathcal{Z} = \big( x_0,t_0; x_1,t_1; x_2,t_2; \cdots; x_{n-1}, t_{n-1}; x, t \big)$ with $x_i \ne 0, 1$ for all $i=0,1,2,\cdots,n-1$, the probability density of this specific path can be given by the product of short-term transition probability densities of consecutive steps. This yields $q(\mathcal{Z})=\prod \limits_{i=1}^n p(x_i,t_i; x_{i-1}, t_{i-1})$, where $p(x_i,t_i;x_{i-1},t_{i-1})$ represents the transition probability from $x_{i-1}$ at time $t_{i-1}$ to $x_i$ at time $t_i$. Consequently, by considering all possible evolutionary paths from $x_0$ to $x$, $p(x,t;x_0)$ can be formulated as the following path integral \cite{Yu1997}:
\begin{align}
p(x,t;x_0) &= \lim_{n \rightarrow \infty} \int_{R} \int_{R} \cdots \int_{R} q(\mathcal{Z}) dx_1 dx_2 \cdots dx_{n-1} \ ,
\end{align}
where $R$ represents the range of the strategy space (i.e., $x\in(0,1)$).

For small $\tau$, we have the following estimate for the transition probability \cite{risken1992}:
\begin{align}
p(x,t+\tau; x',t) &= \frac{1}{2 \sqrt{\pi D^{(2)}(x') \tau}} \exp \left [ -\frac{\big [x-x'-D^{(1)}(x') \tau \big ]^2}{4D^{(2)}(x') \tau} \right ] \ .
\end{align}
Thus, by setting $\tau=t_i -t_{i-1}$ for each step of the evolutionary path, we can use Eq. (5) to express the short-term transition probability density $p(x_i,t_i;x_{i-1},t_{i-1})$. This leads to the following path integral formulation \cite{feynmanhibbs1965, minlan2023} (see Sect. C of the appendix):
\begin{align}
p(x,t;x_0) &= e^{\frac{1}{2} \int_{x_0}^x \frac{D^{(1)}(y)}{D^{(2)}(y)} dy} \int_{(0,x_0)}^{(t,x)}  \frac{e^{- \int_0^t \frac{\dot{x}(s)^2 + (D^{(1)}(x(s)) )^2}{4 D^{(2)}(x(s))} ds}}{ \lim_{n \rightarrow \infty} \prod_{i=1}^n 2 \sqrt{\pi D^{(2)}(x_{i-1}) \tau}} \ \mathfrak{D} x(t) \ ,
\end{align}
where $\dot{x}(s)$ is a shorthand notation for the derivative $dx(s)/ds$. Under weak selection, defined by a small but nonzero $w$ \cite{nowak2004, nowak2006, zheng2011}, we obtain the approximations $D^{(1)}(x)=x(1-x)w(\phi_A(x)-\phi_B(x))$ and $D^{(2)}(x)=x(1-x)/N$. Thus, the probability density for strategy $A$, $p(x,t;x_0)$, can be approximated as (see Sect. C of the appendix):
\begin{align}
p(x,t;x_0) &\approx \exp \Bigg [ \frac{Nw}{2} \int_{x_0}^x \big (\phi_A(y) -\phi_B(y) \big ) dy \Bigg ] \nonumber \\
&\quad \quad \times \int_{(0,x_0)}^{(t,x)} \frac{e^{ - \frac{N}{4} \int^t_0 \left( \frac{\dot{x}(s)^2}{x(s)(1-x(s))} +\sigma_f^2(x(s)) \right ) \ ds}}{\lim_{n \rightarrow \infty} \prod_{i=1}^n 2 \sqrt{\pi x_{i-1} (1-x_{i-1}) \tau /N}} \ \mathfrak{D} x(t) \ .
\end{align}
In the context of population genetics, the term $\exp \Bigg [ \frac{Nw}{2} \int_{x_0}^x \big (\phi_A(y) -\phi_B(y) \big ) dy \Bigg ]$ represents the cumulative effect of directional selection, while the term containing the path integral measure $\mathfrak{D}x(t)$ captures the combined effects of genetic drift ${\dot{x}}^2/(x(1-x))$ and fitness variance $\sigma_f^2(x)$ \cite{Schraiber2014} (see Sect. C of the appendix). Consequently, under neutrality (i.e., $w=0$), $\tilde{p}(x,t;x_0)$ can be expressed as a path integral reflecting solely the influence of genetic drift \cite{wang2023},
\begin{align}
\tilde{p}(x,t;x_0) &= \int_{(0,x_0)}^{(t,x)} \frac{e^{ - \frac{N}{4} \int^t_0 \frac{\dot{x}(s)^2}{x(s)(1-x(s))} \ ds}}{\lim_{n \rightarrow \infty} \prod_{i=1}^n 2 \sqrt{\pi x_{i-1} (1-x_{i-1}) \tau /N}} \ \mathfrak{D} x(t) \ .
\end{align}

Under weak selection, as $w$ is close to zero, $w^{2}$ becomes infinitesimal. Consequently, the term $\int_0^t\sigma_f^2(x) ds=w^2 \int_0^{t} x(1-x) \left ( \phi_A(x) -\phi_B(x) \right )^{2}ds$ in Eq. (7) becomes negligible compared to $\int_0^{t} \frac{\dot{x}(s)^2}{x(s)(1-x(s))}ds$. Thus, $p(x,t;x_0)$ can be re-approximated as:
\begin{align}
p(x,t; x_0) &\approx \mathcal{N} e^{\frac{Nw}{2} \int_{x_0}^x \left ( \phi_A(y) -\phi_B(y) \right ) dy } \ \tilde{p}(x,t;x_0) \ ,
\end{align}
where $\mathcal{N}$ is the normalization constant such that $\int_0^1 p(x,t;x_0)=1$. This expression (9) serves as the path integral solution of Eq. (2) under weak selection. It indicates that the stochastic process of evolution under weak selection and genetic drift, $p(x,t;x_0)$, can be viewed as a modification of the corresponding genetic drift process, $\tilde{p}(x,t;x_0)$, by the cumulative effect of directional selection $e^{\frac{Nw}{2} \int_{x_0}^x \left ( \phi_A(y) -\phi_B(y) \right ) dy }$.

Under weak selection, the ratio of $p(x,t;x_0)$ to $\tilde{p}(x,t;x_0)$, denoted $\gamma(x,x_0)$, depicts a path characteristic that is independent of time $t$:
\begin{align}
\gamma(x,x_0) &\equiv \frac{p(x,t;x_0)}{\tilde{p}(x,t;x_0)} = \mathcal{N} e^{\frac{Nw}{2} \int_{x_0}^x \left ( \phi_A(y) -\phi_B(y) \right ) dy } \ .
\end{align}
The ratio effectively captures the cumulative effect of directional selection on the frequency of phenotype $A$ as it moves from $x_0$ to $x$ in the strategy space, irrespective of the time taken to traverse this path. If $\gamma(x,x_0)>1$, the evolutionary path from $x_0$ to $x$ is more likely to be driven by the cumulative effect of natural selection and genetic drift combined, rather than by drift alone. Conversely, if $\gamma(x,x_0)<1$, this path is less favored by the cumulative effect of natural selection and genetic drift than by drift alone. When $\gamma(x,x_0)=1$, the cumulative effect of natural selection becomes indistinguishable from the effect of genetic drift. This interpretation aligns with a heuristic proposed by Kimura for diploid loci, which describes the balance between natural selection and genetic drift \cite{kimura1983}. For any stochastic evolutionary process defined by the payoff matrix and selection intensity in Eq. (2), all realized evolutionary paths from $x_0$ to $x$ can be benchmarked against this time-invariant path characteristic curve, $\gamma(x,x_0)$ as a function of $x$. Thus, the ratio $\gamma(x,x_0)$ encapsulates a fundamental time invariance of stochastic evolutionary dynamics: the cumulative effect of directional selection relative to genetic drift, valid under weak selection over short time scales where the underlying approximations hold.

\textbf{\emph{Expected path characteristic for evolutionary progression.}}
Although the evolutionary process from $x_0$ to $x$ can be characterized by the time-invariant path characteristic $\gamma(x,x_0)$ that reflects the cumulative effect of directional selection from $x_0$ to $x$, illustrated by red curves in \textbf{Fig. 1}, the probability density $p(x,t;x_0)$ of an evolutionary path traversing from $x_0$ to $x$ by time $t$ is time dependent by definition. For a given initial state $x_0$, the expected path characteristic at time $t$, $\left< \gamma(x,x_0) \right > = \int_0^1 \gamma(x,x_0) \ p(x,t;x_0) dx$, represents the average of these cumulative effects over all realized evolutionary paths that reach $x$ by time $t$. Because $\gamma(x,x_0)$ is invariant over time (i.e., $d \gamma/dt=0$), the time derivative of $\left <\gamma(x,x_0) \right >$〉 becomes:
\begin{align}
\frac{d \left < \gamma(x,x_0) \right >}{dt} &= \int_0^1 \gamma(x,x_0) \ \frac{\partial p(x,t;x_0)}{\partial t} dx \ .
\end{align}
As aforementioned, under weak selection, the Fokker-Planck equation (Eq. 2) can be approximated by setting $D^{(1)}(x)=x(1-x)w(\phi_A(x)-\phi_B(x))$ and $D^{(2)}(x)=x(1-x)/N$ (see Sect. C of the appendix). With these approximations and the same boundary conditions for the neutral theory of evolution solely by genetic drift \cite{kimura1955}, $p(x,t;x_0) |_{x=0, \ x=1}=0$ and $\frac{\partial p(x,t;x_0)}{\partial x} \big |_{x=0, \ x=1}=0$, we can express the time derivative of $\left <\gamma(x,x_0) \right >$ under weak selection (small but positive $w$) as the following (see Sect. D of the appendix),
\begin{align}
\frac{d \left < \gamma(x,x_0) \right >}{dt} &= \int_0^1 \gamma(x,x_0) \ (\frac{3Nw^2}{4} als(M)^2 (x-x^* )^2+\frac{w}{2} als(M))x(1-x)p(x,t;x_0) dx \ .
\end{align}
Note that $\alpha \equiv \int_0^1 \gamma(x,x_0)(x-x^*)^2 x(1-x)p(x,t;x_0)dx$ and $\beta \equiv \int_0^1 \gamma(x,x_0)x(1-x)p(x,t;x_0)dx$ are the expected values of two non-negative functions that are zero only at finitely many $x$ values (i.e., when $x=0,1,x^*$). By standard properties of the Lebesgue integral \cite{Folland1999}, it follows that both integrals are strictly positive, i.e., $\alpha>0$ and $\beta>0$. Consequently, Eq. (12) can be further simplified to:
\begin{align}
\frac{d \left<\gamma(x,x_0)\right>}{dt}=w(\frac{3Nw}{4} als(M)^2 \alpha+\frac{1}{2} als(M)\beta) \ .
\end{align}
Under neutrality ($w=0$), we have $\frac{d\left<\gamma(x,x_0)\right>}{dt}=0$. Under weak selection (small but $w>0$), for games with $als(M)=0$ (nonneutral but with $\sigma_f^2(x)=0$ for all $x\in(0,1)$), we have $\frac{d\left<\gamma(x,x_0)\right>}{dt}=0$; otherwise, for games with $als(M)\ne0$ (nonneutral but with $\sigma_f^2 (x)\ge0$ almost everywhere, except at $x=0,1,x^*$ when $\sigma_f^2 (x)=0$), we have $\frac{d\left<\gamma(x,x_0)\right>}{dt}>0$ provided that $Nw>\frac{2\beta}{3\mid als(M)\mid \alpha}$, which is satisfied for a large population ($Nw\gg1$; see \textbf{Fig. 2}).

Although the cumulative effect of directional selection from state $x_0$ to $x$, i.e., $\gamma(x,x_0)$, is time invariant and fixed, the evolutionary process over time $t$ increasingly favors the paths with a greater path characteristic $\gamma(x,x_0)$ and as a result increases the expected path characteristic $\left <\gamma(x,x_0)\right >$. The cumulative effect of directional selection on expected phenotypic frequency becomes increasingly discernable over time, provided there is a standing fitness variance ($\sigma_f^2 (x)>0$ almost everywhere, for small $w$ and large $N$). The ever-increasing tendency of the average path characteristic of realized evolutionary paths, $\left <\gamma(x,x_0)\right >$, serves as an indicator of evolutionary progression in stochastic evolutionary games with a large population under weak selection and genetic drift.

The time invariance of $\gamma(x,x_0)$ and the non-decreasing property of $\left<\gamma(x,x_0)\right>$ also hold in frequency-independent models. For instance, consider a frequency-independent model of two alleles, $a$ and $A$, with the frequency of $A$ denoted by $x$. The fitnesses of the genotypes are given by $f_{AA}=1+2s,f_{Aa}=1+s,f_{aa}=1$, and the population size is fixed at $N$. In the Wright–Fisher diffusion process \cite{ewens2004, song2012}, with $s=\mathcal{O}(N^{-1})$, the two terms of the Fokker-Planck equation for allele $A$ are: $D^{(1)} (x)=2Nsx(1-x)$ and $D^{(2)} (x)=\frac{1}{2} x(1-x)$, which lead to $\gamma(x,x_0)=e^{2Ns(x-x_0)}$ and $\left<\gamma(x,x_0)\right>=6N^2 \int_0^1 \gamma(x,x_0) \sigma_f^2 (x)p(x,t;x_0)dx\ge0$. This confirms the invariance and non-decreasing expectation of the evolutionary patch characteristic $\gamma(x,x_0)$.

\textbf{\emph{Numerical simulations.}}
The time invariance of the ratio $\gamma(x,x_0)$, derived from the path integral formulation, is validated using stochastic simulations based on the Moran process, a standard model of evolution in a finite population under natural selection and genetic drift. We consider three games \cite{maynardsmith1982}: Example 1 represents the Hawk-Dove game, with the payoff matrix $A_1=(9,18;27,9)$ and the evolutionary singularity $x^*=1/3$; Example 2 represents the Stag-Hunt game, with the payoff matrix $A_2=(20,1;1,20)$ and the singularity $x^*=1/2$; and Example 3 represents the Prisoner’s Dilemma, with the payoff matrix $A_3=(3,0;5,1)$ and $x^*=-1\notin(0,1)$. In Example 1, neither of pure strategy $A$ or $B$ is an evolutionarily stable strategy (ESS), while the evolutionary singularity is the ESS. In Example 2, both pure strategies of $A$ and $B$ are ESS, and the evolutionary singularity is unstable. In Example 3, strategy $B$ (defection) is the ESS \cite{maynardsmith1982, hofbauersigmund1998}. \textbf{Fig. A2} illustrates the evolution of the frequency of strategy $A$ over time.

With the Moran process, the evolutionary process of strategy $A$ from $x_0$ was repeated 40,000 times under each payoff matrix and selection intensity ($w=0.01, 0$), with the value $x(t)$ recorded at time steps $t=$1000, 3000, and 6000. The state space $[0,1]$ of $x$ was divided into hundredth-width bins (i.e., $[0,0.01],\cdots,[0.99,1]$), and we counted the number of repeated runs in each bin. For each bin at a specific time step, let $N_1$ ($w=0.01$) and $N_0$ ($w=0$) be the numbers of repeated runs fallen in the bin. The sampling ratio $N_1/N_0$ was then computed for each bin (asterisks in \textbf{Fig. 1}), which showed a strong agreement with the theoretical formula of $\gamma(x,x_0)$ (Eq. (10), as the red curve). The expected ratio $\left<\gamma(x,x_0)\right>$ and the mean fitness $\bar{f}(x)$ were also calculated based on the sampling counts of 20,000 repeated runs in the hundredth-width bins (\textbf{Fig. 2}). Evidently, the mean fitness is not always non-decreasing under natural selection ($0<w<1$), indicating that Fisher’s theorem does not hold in stochastic evolutionary games under frequency-dependent selection. In contrast, the expectation of the path characteristic $\left<\gamma(x,x_0)\right>$ monotonically increases over time, indicating that evolutionary paths initiated at $x_0$ increasingly deviate from those governed solely by genetic drift, moving toward $x$ values that, on average, increase the invariant ratio $\gamma(x,x_0)$.
\begin{figure}[H]
    \begin{center}
    \includegraphics[width=1\linewidth]{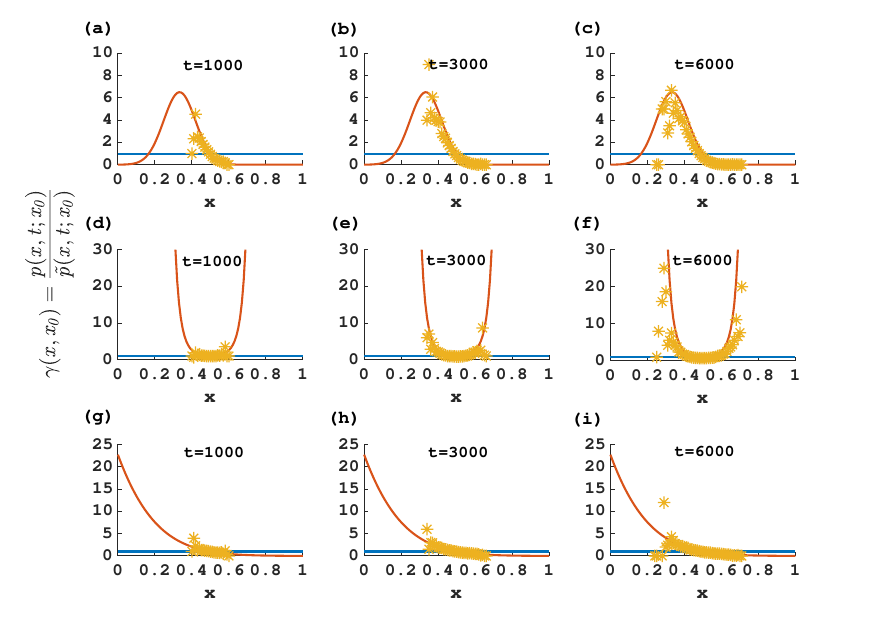}
    \end{center}
    \caption{\textbf{Time invariance of the evolutionary path characteristic $\gamma(x,x_0)$.} Stochastic simulations of the Moran process were performed using three payoff matrices $A_1$, $A_2$ and $A_3$. The population size was set to $N=1000$, with an initial phenotypic frequency $x_0=0.5$. Panels (a-c) show results for $A_1$ at time steps $t=$1000, 3000, and 6000, respectively; panels (d-f) show results for $A_2$; and panels (g-i) show results for $A_3$. The red curves represent $\gamma(x,x_0)$ as described by Eq. (10), while the blue lines represent the path characteristic under neutrality (i.e., $\gamma(x,x_0)=1$ when $w=0$). Asterisks denote the sampling ratio $N_1/N_0$ in each bin of width 0.01 between 0 and 1, calculated from 40,000 repeated simulations, where $N_1$ and $N_0$ represent the number of simulations under weak selection ($w=0.01$) and neutrality ($w=0$), respectively, that reached the bin at the given time point.}
    \label{Fig1}
\end{figure}

\begin{figure}[H]
    \begin{center}
    \includegraphics[width=1\linewidth]{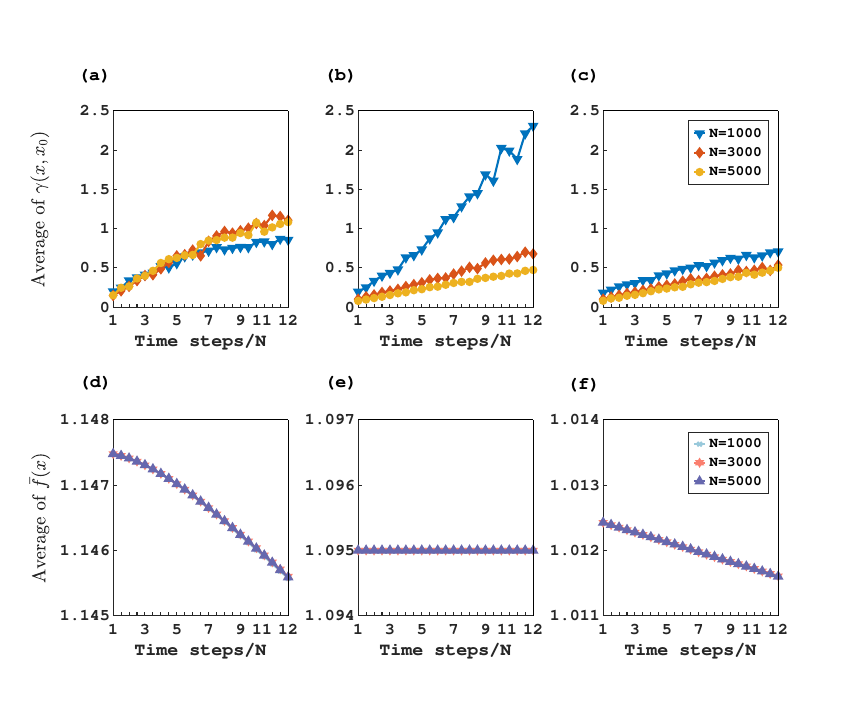}
    \end{center}
    \caption{\textbf{Time evolution of the average path characteristic $\left<\gamma(x,x_0)\right>$ and the average mean fitness $\bar{f}(x)$.} Using the same payoff matrices ($A_1$, $A_2$ and $A_3$, for the columns, respectively) and stochastic simulations as in \textbf{Fig. 1}, each set of simulations was repeated 20,000 times to compute sampling expectations of the path characteristic $\left<\gamma(x,x_0)\right>$ (panels a-c) and mean fitness $\bar{f}(x)$ (panels d-f) at 23 distinct time points. Results are shown for selection intensity $w=0.01$, initial frequency $x_0=0.5$, and population sizes $N=$1000, 3000, and 5000, respectively.}
    \label{Fig2}
\end{figure}

\textbf{\emph{Discussion.}} In finite populations, natural selection and genetic drift act simultaneously in driving the evolution. Under weak selection and drift, based on the diffusion approximation and path integral formulation, an evolutionary path characteristic is identified to be time-independent, and the expectation of this characteristic is also shown to be non-decreasing over time. The evolutionary process of the phenotype frequency is characterized by $\gamma(x,x_0)$, defined by the ratio of the probability density $p(x,t;x_0)$ to the probability density under neutrality $\tilde{p}(x,t;x_0)$, which represents the cumulative effect of directional selection from $x_0$ to $x$. Under neutrality (without selection intensity, $w=0$), the fitnesses $f_A (x)$ and $f_B (x)$ of the two phenotypes are equal, with the ratio $\gamma(x,x_0)=1$, and the mean fitness $\bar{f}(x)$ remains constant. Under weak selection (small but positive $w$), this ratio is time invariant, and equals the sampling ratio of the number of realized evolutionary paths reaching $x$ from $x_0$ over finite time in repeated simulations to the number of realized evolutionary paths under neutrality (\textbf{Fig. 1}). This confirms that the effect of natural selection on phenotypic evolution over evolutionary paths with fixed initial and final states ($x_0$ and $x$) is deterministic and time-invariant \cite{Schraiber2014}. Relative to genetic drift, natural selection increasingly favors the evolutionary paths with $\gamma(x,x_0)>1$ but disfavors those with $\gamma(x,x_0)<1$. This benchmark, time-invariant role of $\gamma(x,x_0)$ suggests that it can be regarded as the path characteristic for stochastic evolutionary dynamics.

With the path integral formulation, the expected characteristic of realized evolutionary paths, $\left<\gamma(x,x_0)\right>$, increases monotonically with time when fitness variance is present. The non-decreasing nature of the expected path characteristic $\left<\gamma(x,x_0)\right>$ also reflects that the average phenotypic frequency $\bar{x}(t)$ at time $t$ over repeated runs shifts toward enhancing $\left<\gamma(x,x_0)\right>$, indicating the directional selection. This ever-increasing property is generic for stochastic evolutionary dynamics but subject to certain conditions, namely, weak selection (small but positive $w$), a large population size (such that $Nw\gg1$), and fitness differences between phenotypes (to ensure fitness variance $\sigma_f^2 (x)>0$ almost everywhere).

Fisher’s fundamental theorem describes the role of natural selection in deterministic frequency-independent evolution, where selection enhances mean fitness, although this role is often smudged by "environmental deterioration" \cite{Okasha2008}. Strict simplification and assumptions are required to ensure $d\bar{f}/dt=\sigma_f^2$ \cite{ewens2024}. Our model broadened the evolutionary process by considering genetic drift and frequency-dependent fitness under weak selection, which leads to the discovery of the time-invariant path characteristic and its non-decreasing expectation. Consequently, at least for stochastic evolutionary games with two strategies, Fisher’s fundamental theorem can be rephrased in terms of $\left<\gamma(x,x_0)\right>$ rather than mean fitness. As the biological setting behind our model is much more realistic compared to those underlying Fisher’s theorem, we think the expected path characteristic can serve as an alternative metric for population-level performance and demonstrate the general principle of evolutionary progression in the stochastic games under natural selection and drift.

Different from the traditional frequency-dependent evolutionary games focusing on the ESS of adaptive strategy $x$ over long-term evolution \cite{traulsen2005, traulsen2006}, the invariance of $\gamma(x,x_0)$ and the non-decreasing property of $\left<\gamma(x,x_0)\right>$ are both conclusions for a finite timeframe (i.e., before phenotypic frequencies can reach the absorption or fixation state). Over long-term evolution ($t \rightarrow \infty $), it is evident that $\tilde{p}(x,t;x_0)\big |_{x\in(0,1)}\to 0$ for evolutionary games driven solely by genetic drift, and $p(x,t;x_0) \big |_{x\in(0,1)} \to 0$ for frequency-dependent stochastic games without a stable internal ESS, while the probability densities of the absorbing states (at the boundaries of the strategy space, i.e., $x=0,1$) approach unity due to the regularity condition, which contradicts the boundary conditions of the model (i.e., $p(x,t;x_0 ) \big |_{x=0, \ x=1}=0$). We thus only consider the Fokker-Planck equation of evolution and the path integral formulation to be valid descriptions of short-term evolutionary paths in frequency-dependent stochastic evolutionary games.

Empirical tests of our theory are challenging but possible. Since $\gamma(x,x_0)$ represents the ratio of probability density under weak selection and genetic drift ($p(x,t;x_0)$) to that under drift alone ($\tilde{p}(x,t;x_0)$), it offers a quantitative metric for distinguishing the roles of selection and drift in evolution. This insight provides clues to experimental designs aimed at measuring changes in biological characteristics over time under both selective and neutral conditions. For example, one could examine the transfer rates of different plasmid types in bacteria exposed to varying antibiotic treatments, environmental conditions, or cell densities (e.g., \cite{Michaelis2023, wang2024}). Such experiments would help differentiate the relative contributions of standing selective pressures versus stochastic drift, thereby providing empirical validation of the theoretical framework. More broadly, $\left<\gamma(x,x_0)\right>$ could serve as a novel indicator for assessing evolutionary performance and directional progression in systems governed by stochastic evolutionary dynamics.

\renewcommand{\theequation}{A\arabic{equation}}
\setcounter{equation}{0}

\renewcommand{\thefigure}{A\arabic{figure}}
\setcounter{figure}{0}

\section*{Appendix}
\textbf{\emph{Section A. Time derivative of mean fitness, $\bar{f}(x)$.}}
In the basic model of a two-phenotype evolutionary game, under the assumptions of weak selection in a large population, the time evolution of the phenotypic frequency $x$ can be approximated as (Traulsen et al. 2005):
\begin{align}
\frac{dx}{dt}=wx(1-x)(\phi_A(x)-\phi_B(x)) \ .
\end{align}
Since that $f_A(x)-f_B(x)=w(\phi_A(x)-\phi_B(x))$, the time derivative of the mean fitness $\bar{f}(x)$ can be written as:
\begin{align}
\frac{d \bar{f}(x)}{d t} & =\frac{d(x f_{A}(x)+(1-x) f_{B}(x))}{d t} \nonumber \\
& =(f_{A}(x)-f_{B}(x)) \frac{d x}{d t}+x \frac{d f_{A}(x)}{d t}+(1-x) \frac{d f_{B}(x)}{d t} \nonumber \\
& =w^{2} x(1-x)(\phi_{A}(x)-\phi_{B}(x))^{2}+x \frac{d f_{A}(x)}{d t}+(1-x) \frac{d f_{B}(x)}{d t} \nonumber \\
& =\sigma_{f}^{2}(x)+x \frac{d f_{A}(x)}{d t}+(1-x) \frac{d f_{B}(x)}{d t} \ .
\end{align}
When the fitnesses $f_A(x)=f_B(x)=1$ are constant (i.e., when selection intensity $w=0$), the mean fitness of the population does not decrease, $\frac{d\bar{f}(x)}{dt}=0$. However, when the fitness values vary over time, the mean fitness of the population typically decreases, as shown in \textbf{Fig. A1} below.

\textbf{\emph{Section B. Derivation of the Fokker-Planck equation, Eq. (2).}}
In the basic model presented in the main text, when the time step of the Moran process is taken as $1/N$, we have:
\begin{align}
& \quad p(x, t+1/N ; x_{0})-p(x, t ; x_{0}) \nonumber \\
& =p(x-1/N, t ; x_{0}) \pi^{+}(x-1/N)+p(x+1/N, t ; x_{0}) \pi^{-}(x+1 / N)-p(x, t ; x_{0}) \pi^{-}(x) \nonumber \\
& -p(x, t ; x_{0}) \pi^{+}(x) \ .
\end{align}
For large $N$, the probability densities $p(x,t+1/N;x_0)$ and $p(x\pm 1/N,t;x_0)$, as well as the transition probabilities, $\pi^\pm(x\mp 1/N)$, can be approximated by their Taylor series expansions around $x$ and $t$:
\begin{align}
p(x, t+1/N ; x_{0}) & \approx p(x, t ; x_{0})+\frac{\partial}{\partial t} p(x, t ; x_{0}) \frac{1}{N} \ , \nonumber \\
p(x \pm 1 / N, t ; x_{0})& \approx p(x, t ; x_{0}) \pm \frac{\partial}{\partial x} p(x, t ; x_{0}) \frac{1}{N}+\frac{\partial^{2}}{\partial x^{2}} p(x, t ; x_{0}) \frac{1}{2 N^{2}} \ ,  \\
\pi^{ \pm}(x \mp 1 / N) & \approx \pi^{ \pm}(x) \mp \frac{\partial}{\partial x} \pi^{ \pm}(x) \frac{1}{N}+\frac{\partial^{2}}{\partial x^{2}} \pi^{ \pm}(x) \frac{1}{2 N^{2}} \ . \nonumber
\end{align}
Using the first (linear) expansion of A4, the left-hand side of Eq. (A3) becomes
\begin{align}
p(x,t+1/N;x_0)-p(x,t;x_0)\approx\frac{\partial}{\partial t}p(x,t;x_0)\frac{1}{N} \ .
\end{align}
Using the other two expansions of A4, the right-hand side of Eq. (A3) simplifies to:
\begin{align}
p(x, t ; x_{0}) \Bigg[\frac{\partial(\pi^{-}(x)-\pi^{+}(x))}{\partial x} \frac{1}{N} \Bigg]+(\pi^{-}(x)-\pi^{+}(x)) \frac{\partial p(x, t ; x_{0}}{\partial x} \frac{1}{N}+\frac{\partial^{2}(\pi^{-}(x)+\pi^{+}(x))}{\partial x} \frac{1}{2 N^{2}} \ .
\end{align}
Thus, the Fokker-Planck equation of $p(x,t;x_0)$ is given by Eq. (2), where
\begin{align}
D^{(1)}(x) & =\pi^{+}(x)-\pi^{-}(x)=\frac{x(1-x) w(\phi_{A}(x)-\phi_{B}(x))}{(1-w)+w \bar{\phi}(x)} \ , \nonumber \\
D^{(2)}(x) & =\frac{\pi^{+}(x)+\pi^{-}(x)}{2 N}=\frac{x(1-x)[2(1-w)+w(\phi_{A}(x)+\phi_{B}(x))]}{2N((1-w)+w \bar{\phi}(x))} \ .
\end{align}

\textbf{\emph{Section C. Derivation of the path integral formulation in Eq. (6) and Eq. (7).}}
Using Eqs. (4) and Eq. (5), along with the property of multiplication (i.e., $\prod_{i=1}^{n}f(x_i)g(x_i)=\prod_{i=1}^{n}f(x_i)\prod_{i=1}^{n}g(x_i)$), $p(x,t;x_0)$ can be expressed as:
\begin{align}
p(x,t;x_0) &= \lim_{n \rightarrow \infty} \int_{R} \int_{R} \cdots \int_{R} \prod_{i=1}^n \frac{1}{2 \sqrt{\pi D^{(2)}(x_{i-1}) \tau}} \nonumber \\
&\quad \quad \quad \quad \times \exp \left [ -\frac{\big [x_i-x_{i-1}-D^{(1)}(x_{i-1}) \tau \big ]^2}{4D^{(2)}(x_{i-1}) \tau} \right ] dx_1 dx_2 \cdots dx_{n-1} \nonumber \\
&= \lim_{n \rightarrow \infty} \int_{R} \int_{R} \cdots \int_{R} \prod_{i=1}^n \frac{1}{2 \sqrt{\pi D^{(2)}(x_{i-1}) \tau}} \nonumber \\
&\quad \quad \quad \quad \times \prod_{i=1}^n \exp \left [ -\frac{\big [x_i-x_{i-1}-D^{(1)}(x_{i-1}) \tau \big ]^2}{4D^{(2)}(x_{i-1}) \tau} \right ] dx_1 dx_2 \cdots dx_{n-1}\ .
\end{align}
Note that
\begin{align}
&\quad \prod_{i=1}^n \exp \Bigg [ - \frac{\big(x_i-x_{i-1}-D^{(1)}(x_{i-1}) \tau \big)^2}{4D^{(2)}(x_{i-1}) \tau} \Bigg ] \nonumber \\
&= \prod_{i=1}^n \exp \Bigg [ - \frac{(x_i -x_{i-1})^2 -2(x_i-x_{i-1}) D^{(1)}(x_{i-1}) \tau +\big(D^{(1)}(x_{i-1}) \big)^2 \tau^2}{4D^{(2)}(x_{i-1}) \tau} \Bigg] \ ,
\end{align}
where the first component of the exponential term on the right-hand side of Eq. (A9) can be rewritten as:
\begin{align}
\prod_{i=1}^n \exp \Bigg [- \frac{(x_i-x_{i-1})^2}{4D^{(2)}(x_{i-1})\tau} \Bigg ] &= \exp \Bigg [ -\sum_{i=1}^n \left (\frac{x_i-x_{i-1}}{\tau} \right )^2 \frac{\tau}{4D^{(2)}(x_{i-1})} \Bigg] \nonumber \\
& \stackrel{n \rightarrow \infty}{\Longrightarrow} \exp \Bigg [ - \int_0^t \frac{\dot{x}(s)^2}{4 D^{(2)}(x(s))} ds \Bigg ] \ ,
\end{align}
and the rest of the exponential term can be rewritten as:
\begin{align}
&\quad \prod_{i=1}^n \exp \Bigg [ \frac{2(x_i-x_{i-1}) D^{(1)}(x_{i-1}) -\big(D^{(1)}(x_{i-1})\big)^2 \tau}{4D^{(2)}(x_{i-1})} \Bigg ] \nonumber \\
&= \exp \left [ \tau \sum_{i=1}^n \frac{2D^{(1)}(x_{i-1}) \frac{x_i -x_{i-1}}{\tau} -\big(D^{(1)}(x_{i-1})\big)^2}{4D^{(2)}(x_{i-1})} \right ] \nonumber \\
& \stackrel{n \rightarrow \infty}{\Longrightarrow} \exp \left [ \int_0^t \frac{2D^{(1)}(x(s)) \dot{x}(s) -\big(D^{(1)}(x(s))\big)^2}{4D^{(2)}(x(s))} ds \right ] \nonumber \\
&= \exp \left [ \frac{1}{2} \int_{x_0}^{x} \frac{D^{(1)}(y)}{D^{(2)}(y)} dy - \int_0^t \frac{\big(D^{(1)}(x(s)) \big)^2}{4 D^{(2)} (x(s))} ds \right ] \ ,
\end{align}
that is,
\begin{align}
&\quad \prod_{i=1}^n \exp \Bigg [ - \frac{\big(x_i-x_{i-1}-D^{(1)}(x_{i-1}) \tau \big)^2}{4D^{(2)}(x_{i-1}) \tau} \Bigg ] \nonumber \\
& \stackrel{n \rightarrow \infty}{\Longrightarrow} \exp \Bigg [ - \int_0^t \frac{\dot{x}(s)^2}{4 D^{(2)}(x(s))} ds + \frac{1}{2} \int_{x_0}^x \frac{D^{(1)}(y)}{D^{(2)}(y)} dy - \int_0^t \frac{\big(D^{(1)}(x(s)) \big)^2}{4 D^{(2)} (x(s))} ds \Bigg ] \ .
\end{align}
Therefore, Eq. (A8) can be written as
\begin{align}
p(x,t;x_0) &= \lim_{n \rightarrow \infty} \int_{R} \int_{R} \cdots \int_{R}  \frac{1}{\prod_{i=1}^n2 \sqrt{\pi D^{(2)}(x_{i-1}) \tau}} \nonumber \\
&\times \prod_{i=1}^n \exp \Bigg [ - \frac{(x_i -x_{i-1}-D^{(1)}(x_{i-1})\tau)^2}{4D^{(2)}(x_{i-1}) \tau} \Bigg]dx_1 dx_2 \cdots dx_{n-1} \nonumber \\
&= \int_{R} \int_{R} \cdots \int_{R}  \frac{1}{\lim_{n \rightarrow \infty}\prod_{i=1}^n2 \sqrt{\pi D^{(2)}(x_{i-1}) \tau}} \nonumber \\
&\times \lim_{n \rightarrow \infty} \prod_{i=1}^n \exp \Bigg [ - \frac{(x_i -x_{i-1}-D^{(1)}(x_{i-1})\tau)^2}{4D^{(2)}(x_{i-1}) \tau} \Bigg]dx_1 dx_2 \cdots dx_{n-1} \nonumber \\
&= \int_{(0,x_0)}^{(t,x)}  \frac{\exp \Bigg [ - \int_0^t \frac{\dot{x}(s)^2}{4 D^{(2)}(x(s))} ds + \frac{1}{2} \int_{x_0}^x \frac{D^{(1)}(y)}{D^{(2)}(y)} dy - \int_0^t \frac{\big(D^{(1)}(x(s)) \big)^2}{4 D^{(2)} (x(s))} ds \Bigg ]}{\lim_{n \rightarrow \infty} \prod_{i=1}^n 2 \sqrt{\pi D^{(2)}(x_{i-1}) \tau}} \ \mathfrak{D} x(t) \nonumber \\
& =\exp \Bigg [\frac{1}{2} \int_{x_0}^x \frac{D^{(1)}(y)}{D^{(2)}(y)} dy\Bigg ] \int_{(0,x_0)}^{(t,x)}  \frac{\exp \Bigg [ - \int_0^t \frac{\dot{x}(s)^2}{4 D^{(2)}(x(s))} ds - \int_0^t \frac{\big(D^{(1)}(x(s)) \big)^2}{4 D^{(2)} (x(s))} ds \Bigg ]}{\lim_{n \rightarrow \infty} \prod_{i=1}^n 2 \sqrt{\pi D^{(2)}(x_{i-1}) \tau}} \ \mathfrak{D} x(t)\ . \nonumber \\ 
\end{align}
Thus, the path integral formulation of $p(x,t;x_{0})$ is obtained as Eq. (6):
\begin{align}
p(x,t;x_0) &= e^{\frac{1}{2} \int_{x_0}^x \frac{D^{(1)}(y)}{D^{(2)}(y)} dy} \int_{(0,x_0)}^{(t,x)}  \frac{e^{- \int_0^t \frac{\dot{x}(s)^2 + (D^{(1)}(x(s)) )^2}{4 D^{(2)}(x(s))} ds}}{\lim_{n \rightarrow \infty} \prod_{i=1}^n 2 \sqrt{\pi D^{(2)}(x_{i-1}) \tau}} \ \mathfrak{D} x(t) \ .
\end{align}
The path integral contains the following components (Eq. A7 in Section B). First, we have
\begin{align}
\frac{D^{(1)}(x)}{D^{(2)}(x)} &= \frac{2N w \big (\phi_A(x) -\phi_B(x) \big)}{2(1-w) + w \big (\phi_A(x) +\phi_B(x) \big)} \ ,
\end{align}
and therefore,
\begin{align}
\frac{\big(D^{(1)}(x) \big)^2}{D^{(2)}(x)} &= \frac{2N \sigma_f^2(x)}{\big [2(1-w) + w \big(\phi_A(x) +\phi_B(x) \big) \big ] \big [ (1-w) +w \bar{\phi}(x) \big ]} \ ,
\end{align}
where $\sigma_f^2(x)=w^2x(1-x)(\phi_A(x)-\phi_B(x))^2$ denotes the fitness variance of the population. Under weak selection, that is, when $w$ is small but $w\neq0$ (Nowak et al. 2004; Nowak 2006; Zheng et al. 2011), the following approximations hold,
\begin{align}
D^{(1)}(x)\approx x(1-x)w(\phi_A(x)-\phi_B(x)),\quad D^{(2)}(x) &\approx \frac{x(1-x)}{N} \ ,
\end{align}
As a result, we have
\begin{align}
\exp \Bigg [ \frac{1}{2} \int_{x_0}^x \frac{D^{(1)}(y)}{D^{(2)}(y)} dy \Bigg ] &\approx \exp \left [ \frac{N w}{2} \int^x_{x_0} \big( \phi_A(y) -\phi_B(y) \big) dy \right ] \ ,
\end{align}
and
\begin{align}
\exp \Bigg [ - \int_0^t \frac{\big(D^{(1)}(x) \big)^2}{4D^{(2)}(x)} ds \Bigg ] &\approx \exp \Bigg [ -\frac{N}{4} \int^t_0 \sigma_f^2(x) ds \Bigg ] \ .
\end{align}
In accordance with Eq. (6), we then obtain $p(x,t;x_0)$ under weak selection, as expressed in Eq. (7).

\textbf{\emph{Section D. Derivation of the time derivative of $\left< \gamma(x,x_0) \right>$, Eq. (12).}}
According to Eq. (2) and Eq. (A17), the time derivative of $\left< \gamma(x,x_0) \right>$ can be expressed as:
\begin{align}
\frac{d\left\langle\gamma(x, x_{0})\right\rangle}{dt} & =\int_{0}^{1} \gamma(x, x_{0}) \frac{\partial p(x, t ; x_{0})}{\partial t} dx \nonumber\\
&=\int_{0}^{1} \gamma(x, x_{0})\Bigg[-\frac{\partial}{\partial x} D^{(1)}(x)p(x,t;x_0 )+\frac{\partial^{2}}{\partial x^{2}} D^{(2)}(x)p(x,t;x_0)\Bigg]dx \nonumber\\
& \approx-\int_{0}^{1} \gamma(x, x_{0}) \frac{\partial}{\partial x} x(1-x) w(\phi_{A}(x)-\phi_{B}(x)) p(x, t ; x_{0}) dx \nonumber\\
& +\frac{1}{N} \int_{0}^{1} \gamma(x, x_{0}) \frac{\partial^{2}}{\partial x^{2}} x(1-x) p(x, t ; x_{0}) dx \ .
\end{align}
The first term of Eq. (A20) can be simplified as follows:
\begin{align}
-\int_{0}^{1} \gamma(x, x_{0}) \frac{\partial}{\partial x} & x(1-x) w(\phi_{A}(x)-\phi_{B}(x)) p(x, t ; x_{0}) dx \nonumber \\
& =-\int_{0}^{1} \gamma(x, x_{0}) d[x(1-x) w(\phi_{A}(x)-\phi_{B}(x)t) p(x, t ; x_{0})] \nonumber \\
& =-\gamma(x,x_0)[x(1-x)w(\phi_A(x)-\phi_B(x)) p(x,t;x_0 )]\big |_{0}^1 \nonumber \\
& +\int_0^1 d[x(1-x)w(\phi_A(x)-\phi_B(x)) p(x,t;x_0)]d\gamma(x,x_0) \nonumber \\
& =\int_{0}^{1}[x(1-x) w(\phi_{A}(x)-\phi_{B}(x)) p(x, t ; x_{0})] \frac{\partial \gamma(x, x_{0})}{\partial x} dx \nonumber \\
& =\frac{N}{2} \int_{0}^{1} \gamma(x, x_{0}) \sigma_{f}^{2}(x) p(x, t ; x_{0}) dx \ , \nonumber
\end{align}
where $\sigma_{f}^{2}(x)=w^2x(1-x)(\phi_{A}(x)-\phi_{B}(x))^2$ represents the fitness variance. The second term of Eq. (A20) simplifies to:
\begin{align}
\frac{1}{N} \int_{0}^{1}& \gamma(x, x_{0}) \frac{\partial^{2}}{\partial x^{2}} x(1-x) p(x, t ; x_{0}) dx =\frac{1}{N} \int_{0}^{1} \gamma(x, x_{0}) d\Bigg[\frac{\partial}{\partial x} x(1-x) p(x, t ; x_{0})\Bigg] \nonumber \\
& =-\frac{1}{N} \int_{0}^{1} \frac{\partial}{\partial x} x(1-x) p(x, t ; x_{0}) d \gamma(x, x_{0})=\frac{1}{N} \int_{0}^{1} x(1-x) p(x, t ; x_{0}) \frac{\partial^{2}(\gamma(x, x_{0}))}{\partial x} dx \nonumber \\
& =\frac{N}{4} \int_{0}^{1} \gamma(x, x_{0}) \sigma_{f}^{2}(x) p(x, t ; x_{0}) d x-\frac{w((c-d)-(a-b))}{2} \int_{0}^{1} \gamma(x, x_{0}) x(1-x) p(x, t ; x_{0}) dx \ . \nonumber
\end{align}
Using the integration by parts formula, $\int u(x)v'(x)dx=u(x)v(x)-\int u'(x)v(x)dx$, and applying the boundary conditions as specified in the main text, we obtain:
\begin{align}
\frac{d\left< \gamma(x,x_0) \right>}{dt} &\approx \int_{0}^{1} \gamma(x, x_{0})\Bigg(\frac{3N}{4}\sigma_{f}^{2}(x)-\frac{w((c-d)-(a-b))}{2}x(1-x)\Bigg) p(x, t ;x_{0}) dx \ .
\end{align}
We notice that
\begin{align}
\frac{3N}{4}\sigma_{f}^{2}(x)=\frac{3Nw^2}{4}x(1-x)(\phi_A(x)-\phi_B(x))^2=\frac{3Nw^2}{4} x(1-x)als(M)^2 (x-x^* )^2 \nonumber \ ,
\end{align}
and that
\begin{align}
-\frac{w((c-d)-(a-b))}{2}=\frac{w}{2}als(M) \nonumber \ .
\end{align}
Consequently, Eq. (A20) can be written as Eq. (12).

\section*{Supplementary Figures}
\begin{figure}[H]
    \begin{center}
    \includegraphics[width=1\linewidth]{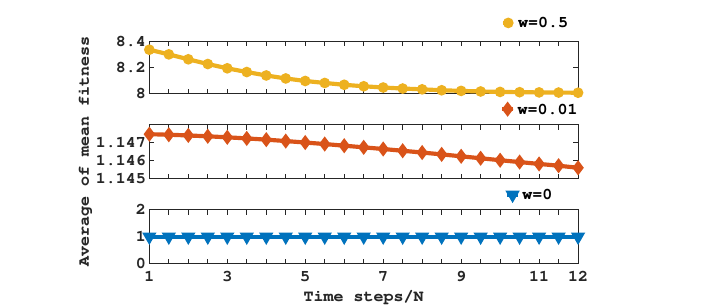}
    \end{center}
    \caption{\textbf{Time evolution of $\bar{f}(x)$.} Stochastic simulations of the Moran process were performed using the payoff matrix $A_1=(9,18;27,9)$, with the mean fitness $\bar{f}(x)$ calculated at 23 distinct time points. The population size was set to $N=1000$, with the initial state $x_0=0.5$. Each plot represents the average of mean fitness from 40,000 repeated stochastic simulations. The yellow, red, and blue lines represent the time evolution of the average of mean fitness under selection intensities $w=0.5,\ 0.01$, and 0, respectively.}
    \label{FigA1}
\end{figure}

\begin{figure}[H]
    \begin{center}
    \includegraphics[width=1\linewidth]{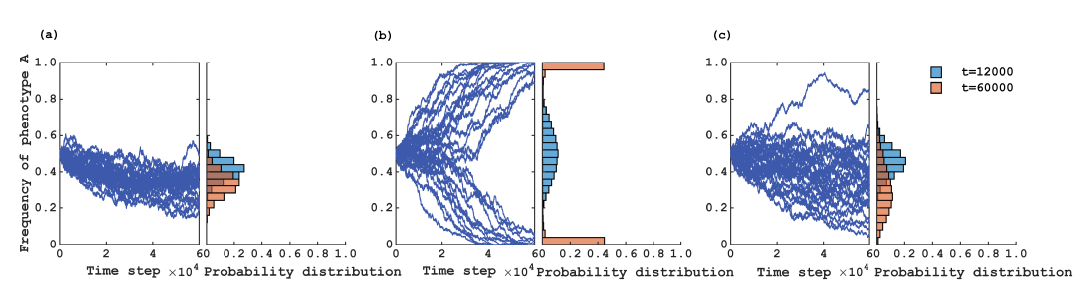}
    \end{center}
    \caption{\textbf{Time evolution of the frequency $x$ of phenotype $A$.} Stochastic simulations of the Moran process were performed based on the payoff matrices $A_1=(9,18;27,9)$ for (a), $A_2=(20,1;1,20)$ for (b), and $A_3=(3,0;5,1)$ for (c). The simulation represents a population of size $N=1000$ and an initial state of $x_0=0.5$ for selection intensity $w=0.01$ over $t=60,000$ time steps. Each plot includes 20,000 repeated simulations, indicated by the blue lines. The blue and orange vertical histograms, with a bin width of 0.04, represent the probability density distributions of the phenotypic frequency $x$ at time steps $t=12,000$ and $t=60,000$, respectively.}
    \label{FigA2}
\end{figure}

\section*{Acknowledgements}

\textbf{Funding:} This work was supported by the NSFC-Yunnan United fund (Grant No. U2102221) and the National Natural Science Foundation of China (Grants No.32071610 and No.31971511). CH also acknowledges the support from the National Research Foundation (No. 89967).\\


\bibliographystyle{unsrt}


\end{document}